\documentclass[10pt]{iopart}
\usepackage{amsfonts}

\begin{document}

\title                  {Brane-world Cosmologies with non-local bulk effects}

\author                 {R J  van den Hoogen, Adam Horne}
\address                {Department of Mathematics, Statistics, and Computer Science,
                        Saint Francis Xavier University, Antigonish, N.S.,
                        B2G 2W5, Canada}
                        
\ead                  {rvandenh@stfx.ca}

\begin{abstract}

It is very common to ignore the non-local bulk effects in the study of brane-world cosmologies using the brane-world approach.  However, we shall illustrate through the use of three different scenarios, that the non-local bulk-effect ${\cal P}_{\mu\nu}$ does indeed have significant impact on both the initial and future behaviour of brane-world cosmologies.

\end{abstract}

%\submitto{\CQG}


\maketitle

%% -------------------------------------------------------------------------
%% --------------            Section 1              ------------------------
%% -------------------------------------------------------------------------

\section{Introduction}

  There is much interest in cosmologies derived from alternative theories of gravity.  String theory, being one of these alternatives has motivated many different viable cosmologies.  In particular there is a class of cosmologies motivated by string theory in which matter fields are restricted to a $3$-dimensional brane embedded in $1+3+d$ dimensions (the bulk), while the gravitational field is free to propagate in the $d$ extra dimensions \cite{Rubakov:1983,Akama:1983,Visser:1985,Squires:1986,Arkani-Hamed:1998,Antoniadis:1998}.  In fact, Randall and Sundrum \cite{Randall:1999a,Randall:1999b} have been successful in recreating conventional Einstein gravity at low enough energies in their second brane-world proposal \cite{Randall:1999b}, here-after labelled RS2.  There has been many studies of cosmologies derived from these brane-world scenarios.  For example, the so-called perfect fluid Friedmann brane-world models were investigated in \cite{Binetruy:2000a,Csaki:1999,Cline:1999,Kaloper:1999,Mukohyama:2000,Binetruy:2000b,Campos:2001a,Campos:2001b} and more recently, people have investigated anisotropic brane-world models \cite{Campos:2001a,Campos:2001b,Campos:2003a,Santos:2001,Maartens:2001b,Toporensky:2001,Coley:2002a,Savchenko:2002,vandenHoogen:2003a,vandenHoogen:2003b,Aguirregabiria:2003a}.  See \cite{Maartens:2003,Langlois:2002a,Padilla:2002,Brax:2003} for reviews of brane-world cosmology. 

There are essentially two approaches in studying brane-world cosmologies.  Firstly, the bulk-based approach \cite{Binetruy:2000a,Csaki:1999,Cline:1999,Binetruy:2000b} in which one first solves the higher dimensional field equations and then attempts to determine whether a lower-dimensional brane-world exists and subsequently studies the cosmological dynamics on the brane. Secondly, the brane-world approach in which one solves the projected field equations on the brane \cite{Shiromizu:2000,Maartens:2000a} and essentially ignores the extension of the solution to the bulk.  Both of these fundamentally different approaches have their benefits as well as their drawbacks (See \cite{Campos:2003a} for additional arguments).

One of the inherent problems of studying brane-world cosmologies using the brane-world approach is that observers on the brane cannot possibly determine all the effects resulting from the bulk.  In particular there exist non-local effects $({\cal U},{\cal Q}_\mu,{\cal P}_{\mu\nu})$ coming from the bulk that affect the local dynamics of observers on the brane \cite{Maartens:2000a}.  It is not at all well understood what the non-local effects from the bulk have on cosmological observations in the brane.  In most papers that study RS2 brane-world cosmologies, one of the following assumptions is often made:
\begin{itemize}
\item ${\cal U}={\cal Q}_\mu={\cal P}_{\mu\nu}=0$,   in which case the bulk is conformally flat and there are no non-local bulk effects affecting the dynamics on the brane.
\item ${\cal U}\not = 0$ but ${\cal Q}_\mu={\cal P}_{\mu\nu}=0$, which are necessary conditions for the existence of a Friedmann Brane-world while the remaining non-zero component of the non-local effect on the brane, ${\cal U}$, behaves as if it were a radiative fluid (assuming it is positive), or can be interpreted holographically as the dual conformal field theory living on the brane \cite{Savonije:2001,Gregory:2002}.
\item ${\cal U}\not = 0$, ${\cal Q}_\mu=0 $ and $\sigma^{\mu\nu}{\cal P}_{\mu\nu}=0$, a common constraint in the investigation of anisotropic but spatially homogeneous brane-worlds, chosen primarily for the ease in analyzing the models in a self consistent manner \cite{Maartens:2001b,Toporensky:2001}.
\end{itemize}
There has been some additional work on determining the effects of a non-zero ${\cal P}_{\mu\nu}$ on the behaviour of brane-world cosmological models \cite{Maartens:2001b,Toporensky:2001,Savchenko:2002,Aguirregabiria:2003a,Barrow:2002b}.  However, for the most part, all the assumptions above are essentially adhoc and not well founded.  Below, we will make an alternative but reasonable assumption on the form of the unknown ${\cal P}_{\mu\nu}$.

A note with regards to notation.   Tilde quantities denote quantities in the five dimensional bulk while non-tilde quantities are quantities on the four-dimensional 3-brane.  The upper case latin indices range from $0$ to $4$ while the lower case greek indices range from $0$ to $3$. Round (square) brackets enclosing indices represent (anti-) symmetrization.

%% -------------------------------------------------------------------------
%% --------------            Section 2              ------------------------
%% -------------------------------------------------------------------------

\section{The Effective Field Equations on the Brane}

We shall investigate brane-world cosmologies using the brane-world approach.  Shiromizu et al. \cite{Shiromizu:2000} have developed an elegant covariant approach to determine the effective four dimensional field equations induced on a single brane embedded in a five dimensional bulk  using the Gauss-Codazzi equations, the so-called Israel junction conditions \cite{Israel:1966} and the usual $Z_2$ symmetry. Maartens \cite{Maartens:2000a} further developed the ideas of Shiromizu et al. by extending earlier work of Ellis and MacCallum \cite{EllisMacCallum:1969,WainwrightEllis:1997} by determining the dynamical equations describing braneworld cosmologies.   The resulting equations are a modification of the standard Einstein Field
equations, with new terms transferring bulk effects onto the
brane. Assuming a five-dimensional bulk spacetime containing a 3-brane,  the field equations are Einstein's equations with a (negative) bulk cosmological constant $\widetilde \Lambda$ and brane energy momentum $T_{AB}$, as a source
$$\widetilde G_{AB}=\widetilde \kappa^2\left[-\widetilde\Lambda \tilde g_{AB} + \delta(\chi)\{-\lambda g_{AB} +T_{AB}\}\right]$$
where $\chi=0$ yields the location of the 3-brane and $\widetilde\kappa^2=8\pi/\widetilde M_p^3$.  Let $g_{\mu \nu}$ be the induced metric on the brane, then the effective field equations on the brane become
\begin{equation}
G_{\mu\nu}=-\Lambda g_{\mu\nu}+\kappa^2 T_{\mu\nu}+\widetilde{\kappa}^4S_{\mu\nu} - {\cal E}_{\mu\nu}\,,
\end{equation}
where
\begin{equation}
\kappa^2=\frac{8\pi}{M_{\rm p}^2}\,,~~\lambda=6{\frac{\kappa^2}{\widetilde\kappa^4}} \,, ~~ \Lambda =
\frac{4\pi}{ \widetilde{M}_{\rm p}^3}\left[\widetilde{\Lambda}+
\left({\frac{4\pi}{3\widetilde{M}_{\rm
p}^{\,3}}}\right)\lambda^2\right]\,.
\end{equation}

The bulk corrections to the Einstein equations on the brane manifest themselves in
two forms: firstly, any matter field on the brane contributes to local quadratic
energy-momentum corrections via the tensor $S_{\mu\nu}$ defined as
\begin{equation}\label{local}
S_{\mu\nu}=\frac{1}{12}T_\alpha{}^\alpha T_{\mu\nu}
-\frac{1}{4}T_{\mu\alpha}T^\alpha{}_\nu+
\frac{1}{24}g_{\mu\nu} \left[3 T_{\alpha\beta}
T^{\alpha\beta}-\left(T_\alpha{}^\alpha\right)^2 \right]\,.
\end{equation}
Secondly, there are nonlocal effects from the free gravitational
field in the bulk, transmitted through a projection of the bulk Weyl tensor onto the brane.  If $n^A$ is the spacelike unit normal vector to the brane then we define
$${\cal E}_{AB}\equiv\widetilde C_{ABCD}n^Cn^D$$
which is both traceless and symmetric and has no components that are orthogonal to the brane.  We then take $$\lim_{\chi \to 0} {\cal E}_{AB} = {\cal E}_{\mu\nu}\delta^{\mu}_A \delta^\nu_B$$ to define the non-local effect of the bulk gravitational field on the brane \cite{Maartens:2000a}.  Given a timelike congruence $u^\mu$, on the brane, the bulk correction, ${\cal E}_{\mu\nu}$ can be decomposed  into an equivalent non-local energy density on the brane, ${\cal U}$, a nonlocal energy-flux on the brane, ${\cal Q}_\mu$, and a non-local anisotropic pressure on the brane, ${\cal P}_{\mu\nu}$, via
\begin{equation}
{\cal E}_{\mu\nu}=-\left(\frac{\widetilde \kappa}{\kappa}\right)^4\left[{\cal U}(u_{\mu}u_{\nu}+\frac{1}{3}h_{\mu\nu})+{\cal P}_{\mu\nu}+ 2{\cal Q}_{(\mu}u_{\nu)}\right]
\end{equation}
where
\begin{eqnarray}
{\cal U} &=& -\left(\frac{\kappa}{\widetilde \kappa}\right)^4{\cal E}_{\mu\nu}u^{\mu}u^{\nu}, \\
{\cal Q}_{\mu} & = & \left(\frac{\kappa}{\widetilde \kappa}\right)^4{\cal E}_{<\mu>\nu}u^{\nu}, \\
{\cal P}_{\mu\nu} &=& -\left(\frac{\kappa}{\widetilde \kappa}\right)^4{\cal E}_{<\mu\nu>}.
\end{eqnarray}
Where $h_{\mu \nu}=g_{\mu \nu}+u_{\mu}u_{\nu}$ projects orthogonal to $u_\mu$ on the brane and the angled brackets denote the projected, symmetric, and trace free part via
$$ V_{<\mu>}=h_{\mu}^{\ \nu}V_{\nu},\qquad\qquad W_{<\mu \nu>}=[h_{(\mu}^{\ \ \alpha}h_{\nu)}^{\ \ \beta}-\frac{1}{3}h^{\alpha \beta}h_{\mu\nu}]W_{\alpha\beta}$$
(See \cite{Maartens:2000a} for complete details.)

Due to the nature of the matter in the bulk (a cosmological constant) and the $Z_2$ symmetry, it can be shown that the brane energy momentum tensor on the brane is separately conserved, that is
\begin{equation}
T^{\mu}_{\nu;\mu}=0\label{cons1}.
\end{equation}
It then follows from the contracted Bianchi identities on the brane that
\begin{equation}
{\cal E}^{\mu}_{\nu;\mu}=\widetilde\kappa^4 S^{\mu}_{\nu;\mu}\label{cons2}.
\end{equation}
In general, equation (\ref{cons2}) yields evolution equations for ${\cal U}$ and ${\cal Q}_\mu$ (see \cite{Maartens:2000a}) for further details). Unfortunately, there are no
evolution equations for the non-local anisotropic pressure, ${\cal P}_{\mu\nu}$.  Therefore, generically the projection of the
5-dimensional field equations onto the brane does not lead to a
closed system of equations.

%% -------------------------------------------------------------------------
%% --------------            Section 3              ------------------------
%% -------------------------------------------------------------------------

\section{Governing Equations}

In order to illustrate the possible effects of ${\cal P}_{\mu\nu}$ on the dynamics of a braneworld cosmological model, we first make some simplifying assumptions.
\begin{itemize}
\item Firstly, it is common to assume through fine tuning of the brane tension $\lambda$ (a la Randall Sundrum) that the effective cosmological constant on the brane is zero.  Here, for illustrative purposes only, we shall also assume that the effective cosmological constant on the brane is zero, that is, $\Lambda = 0$.
\item The simplest geometry for the braneworld, that permits a non-zero ${\cal P}_{\mu\nu}$ is that of a Bianchi type I spacetime geometry.  That is, we will assume a Bianchi I brane with line element of the form 
\begin{equation}
 ds^2=-dt^2+[a_1(t)]^2dx^2+[a_2(t)]^2dy^2+[a_3(t)]^2dz^2.
 \end{equation}
\item  For simplicity, we shall assume that the matter on the brane is in the form of a perfect fluid with four velocity orthogonal to the surfaces of homogeneity (non-tilted).  The energy-momentum tensor for a perfect fluid on the brane is given by
\begin{equation}
T_{\mu\nu}=\rho u_\mu u_\nu+ph_{\mu\nu}\,
\end{equation}
where $u^\mu$ is the fluid 4-velocity and $\rho$ and $p$ are the energy density and isotropic pressure respectively.
\end{itemize}

It then follows that the Hubble expansion scalar, acceleration vector, and anisotropic shear tensor are of the form
\begin{eqnarray}
H &\equiv & \frac{1}{3} u^{\mu}_{\ ;\mu} = \frac{1}{3}\left(\frac{\dot a_1}{a_1}+\frac{\dot a_2}{a_2}+\frac{\dot a_3}{a_3}\right) \nonumber\\
\dot u_{\mu} &\equiv& u_{\mu;\nu}u^{\nu} = 0 \nonumber\\
\sigma_{\mu\nu} & \equiv & u_{(\mu;\nu)} - H h_{\mu\nu}+ \dot u_{(\mu}u_{\nu)} \nonumber\\
&=& \frac{1}{3}{\rm diag} \left( 0, 
2\frac{\dot a_1}{a_1}-\frac{\dot a_2}{a_2}-\frac{\dot a_3}{a_3},
2\frac{\dot a_2}{a_2}-\frac{\dot a_1}{a_1}-\frac{\dot a_3}{a_3},
2\frac{\dot a_3}{a_3}-\frac{\dot a_1}{a_1}-\frac{\dot a_2}{a_2}\right)\nonumber
\end{eqnarray}
where an over-dot represents a derivative with respect to $t$.

With the assumption of a perfect fluid source on the brane, the local bulk corrections, (see equation \ref{local}), to the effective four-dimensional field equations reduces to
\begin{equation}
S_{\mu\nu} =\frac{1}{12} \rho^2 u_\mu u_\nu
+\frac{1}{12}\rho\left(\rho+2 p\right)h_{\mu\nu}\,.
\end{equation}
Both the local and non-local bulk corrections mentioned above can be consolidated into an effective total
energy density, pressure, heat flux and anisotropic stress  as follows. The modified Einstein equations can be rewritten in the standard
Einstein form with a redefined energy-momentum tensor:
\begin{equation}
G_{\mu\nu}=\kappa^2 T^{\rm total}_{\mu\nu}\,, \label{E1}
\end{equation}
where
\begin{equation}
T^{\rm total}_{\mu\nu} \equiv T_{\mu\nu}+\frac{\widetilde{\kappa}^{4}}{\kappa^2}S_{\mu\nu}- \frac{1}{\kappa^2}{\cal E}_{\mu\nu} \label{E2}
\end{equation}
is equivalent to an imperfect fluid energy-momentum tensor
with the total equivalent energy density, pressure, heat flux and anisotropic pressure due to a perfect fluid source on the brane
\begin{eqnarray}
\rho^{\rm total} \label{total1} &=& \rho +\frac{\widetilde{\kappa}^{4}}{\kappa^6}
\Biggl[\frac{\kappa^4}{12}\rho^2+{\cal U}\Biggr] \label{rho},\\
p^{\rm total} \label{total2} &=&  p+\frac{\widetilde{\kappa}^{4}}{\kappa^6}
\Biggl[ \frac{\kappa^4}{12}\rho
\left(\rho+2p \right)+\frac{1}{3}{\cal U}\Biggr], \\
q^{\rm total}_{\mu} &=& \frac{\widetilde{\kappa}^{4}}{\kappa^6} {\cal Q}_\mu, \\
\pi^{\rm total}_{\mu\nu}&=& \frac{\widetilde{\kappa}^{4}}{\kappa^6} {\cal P}_{\mu\nu}\label{pi}.
\end{eqnarray}

The effective Einstein field equations (\ref{E1}-\ref{E2}) yield the following dynamical equations describing the evolution of the Bianchi I brane-world \cite{Maartens:2000a,Maartens:2001a}:
\begin{eqnarray}
\dot H &=& -H^2 -\frac{2}{3}\sigma^2 -\frac{\kappa^2}{6}(\rho^{\rm total}+3p^{\rm total}), \\
\dot \sigma_{\mu\nu} &=& -3H\sigma_{\mu\nu} + \kappa^2 {\pi}^{\rm total}_{\mu\nu}, \\
\kappa^2\rho^{\rm total} &=& 3H^2-\sigma^2 \label{eq3},\\
\kappa^2 {q^{\rm total}_\mu} & = & 0,
\end{eqnarray}
where $\rho^{\rm total}$, $p^{\rm total}$, $q^{\rm total}_\mu$ and $\pi^{\rm total}_{\mu\nu}$ are defined in equations (\ref{rho}-\ref{pi}) and  $$\sigma^2=\frac{1}{2}\sigma_{\mu\nu}\sigma^{\mu\nu}.$$
From equations (\ref{cons1}-\ref{cons2}) (assuming $p=(\gamma-1)\rho$) we obtain the following conservation equations
\begin{eqnarray}
\dot \rho &=& -3 \gamma H \rho,\\
\dot {\cal U} &=& -4H{\cal U} - \sigma_{\mu\nu}{\cal P}^{\mu\nu},\\
h_{\nu}^{\xi}P^{\nu}_{\mu;\xi}&=&0.\label{cons4}
\end{eqnarray}

We define new dependent dynamical variables of the form
\begin{equation}\label{new_vars}
\begin{tabular}{ll}
$\displaystyle \Sigma = \frac{1}{\sqrt{3}}\frac{\sigma}{H},\qquad\qquad$
& $\displaystyle\Omega =\frac{\kappa^2 \rho}{3H^2},$ \\
 $\displaystyle\Omega_{\cal U}= \frac{6}{\kappa^2\lambda}\frac{\cal U}{3H^2},\qquad\qquad$
& $\displaystyle\Omega_{\lambda}=\frac{3H^2}{2\lambda\kappa^2}\left(\frac{\kappa^2\rho}{3H^2}\right)^2,$
\end{tabular}
\end{equation}
and a new independent variable $$\frac{dt}{d\tau}=\frac{1}{H}.$$ 
We also define a dimensionless variable $\cal P$ to represent the non-local bulk corrections arising from the quantity ${\cal P}_{\mu\nu}$
\begin{equation}\label{P}
{{\cal P}}=\frac{6}{\kappa^2\lambda} \frac{\frac{1}{2}\sigma_{\mu\nu}{\cal P}^{\mu\nu}}      {\sqrt{3} H^2\sqrt{\frac{1}{2}\sigma_{\mu\nu}\sigma^{\mu\nu}}}
= \frac{\sqrt{3}}{\kappa^2\lambda} \frac{\sigma_{\mu\nu}{\cal P}^{\mu\nu}}{\sigma H^2}.
\end{equation}
The resulting system of differential equations decouples into an evolution equation for $H$, (which we use to define the deceleration $q$, )
\begin{equation}
 H'=-(q+1)H \label{H-prime},
\end{equation}
so that
\begin{equation}
q\equiv2\Sigma^2+\Omega_{\cal U}+\frac{1}{2}(3\gamma-2)\Omega+\frac{1}{2}(6\gamma-2)\Omega_{\lambda} \label{q},
\end{equation}
and a reduced dynamical system describing the dynamics of the Bianchi I perfect fluid braneworld with non-local bulk effects
\begin{eqnarray}\label{ds2}
\Sigma' & = & \Sigma(q-2)+ {\cal P}, \\
\Omega' & = & \Omega[2q-(3\gamma-2)],\\
\Omega_{\lambda}' &=& \Omega_{\lambda} [2q-(6\gamma -2)],\\
\Omega_{\cal U}' &=& \Omega_{\cal U}(2q-2) -2\Sigma{\cal P},
\end{eqnarray}
where $'$ denotes differentiation with respect to $\tau$. Equation (\ref{eq3}) yields a constraint of the form
\begin{equation}
1=\Sigma^2+\Omega+\Omega_{\lambda}+\Omega_{\cal U}
\label{constraint2},
\end{equation}
that can be used to globally eliminate one of the dynamical variables.

 There is no evolution equation for the quantity ${\cal P}$.  This is not unexpected since observers on the brane are not expected to be able to determine all characteristics of the bulk (see \cite{Maartens:2000a,Maartens:2001a}).  In order to proceed further, we must make some assumption as to the form of ${\cal P}$. In order to obtain a closed system of equations, it would not be unreasonable to expect that ${\cal P}$ is a differentiable function of the remaining dynamical variables, that is,
\begin{equation}
{\cal P} = {\cal P}(\Sigma,\Omega,\Omega_{\lambda},\Omega_{\cal U}).
\end{equation}
which clearly satisfies the conservation equation (\ref{cons4}).
With this assumption, the value of $q$ (see equation \ref{q}) at
the equilibrium points of the dynamical system (\ref{ds2}) is
constant.  Provided $q\not =-1$ at the equilibrium point, we are
able to solve equation (\ref{H-prime}) and obtain the solution
$H(t)=H_0 t^{-1}$.  With this, we are now able to find the
solution of the remaining variables, $\rho=\rho_0 t^{-2}$,
$\sigma_{\mu\nu}=(\sigma_{\mu\nu})_0 t^{-1}$, ${\cal U} ={\cal
U}_0 t^{-2}$, and ${\cal P}_{\mu\nu} = ({\cal P}_{\mu\nu})_0
t^{-2}$  where $H_0,\rho_0, (\sigma_{\mu\nu})_0,{\cal U}_0,({\cal
P}_{\mu\nu})_0$ are all constants.  It is worth noting, that the
resulting braneworld spacetime represented by these equilibrium
points is geometrically self-similar
\cite{CarrColey:1999,HsuWainwright:1986}.   (If $q=-1$ at an
equilibrium point, then the resulting solution has
$H,\rho,\sigma,{\cal U},$ all equal to constants and consequently
the resulting spacetime is not self-similar.) The property of
asymptotic self-similarity is one that has been investigated by
many in the context of general relativistic cosmologies, see
\cite{CarrColey:1999,HsuWainwright:1986,vandenHoogen:1994}.  Here
we observe, that our assumption on the form of ${\cal P}$ is a
necessary condition in order for the equilibrium points of the
resulting reduced dynamical system to represent self-similar
cosmological models. Now it is not at all clear what the actual
form of this function should be, but we shall investigate a number
of cases.
\begin{itemize}
 \item ${\cal P} = {\cal P}_0$, a constant:  This case includes the scenarios ${\cal P}_{\mu\nu}=0$ studied by \cite{Binetruy:2000a,Kaloper:1999,Mukohyama:2000,Binetruy:2000b,Campos:2001a,Campos:2001b,Coley:2002a,Savchenko:2002,vandenHoogen:2003a,vandenHoogen:2003b} and ${\cal P}_{\mu\nu}\sigma^{\mu\nu}=0$ \cite{Maartens:2001b,Toporensky:2001} as special subcases of ${\cal P}=0$.
 \item ${\cal P} = {\cal P}_{\Sigma}\Sigma^\alpha$,  $\alpha>0$:  This ansatz guarantees that the necessary condition ${\cal P}(\Sigma,\Omega,\Omega_{\lambda},\Omega_{\cal U})|_{\Sigma=0} =0$ is satisfied for the existence of an isotropic equilibrium point.
 \item ${\cal P} = {\cal P}_{\Omega_{\cal U}}\Omega_{\cal U}^\beta$, $\beta>0$: This ansatz is perhaps more desirable than the others in that both ${\cal P}$ and $\Omega_{\cal U}$ are directly related to the Weyl curvature in the bulk, while all the other dynamical variables are inherently on the brane. This assumption is similar to that used in \cite{Savchenko:2002,Barrow:2002b}.
 \end{itemize}

With an assumption on the form of $\cal P$, the evolution equations for the variables $\vec X=[\Sigma,\Omega,\Omega_\lambda,\Omega_{\cal U}]$ yields a dynamical system of the form $\vec X'=\vec F(\vec X)$ suitable for a qualitative analysis using dynamical systems techniques \cite{WainwrightEllis:1997}.  In short, the procedure requires one to find equilibrium solutions of the dynamical system, that is, points $\vec X=\vec X_0$ such that $\vec F(\vec X_0)=\vec 0$.  If the eigenvalues of the linearization of $\vec F(\vec X)$ at $\vec X=\vec X_0$ have non-zero real parts, then the nonlinear system $\vec X'=\vec F(\vec X)$ behaves qualitatively the same as the linear system $\vec X'=DF(\vec X_0)(\vec X-\vec X_0)$ in a local neighborhood of $\vec X = \vec X_0$.  If the real parts of the eigenvalues of $DF(\vec X_0)$ are all negative (positive), then the equilibrium point is said to be stable to the future (past) as all orbits in a local neighborhood of the equilibrium point have this equilibrium point as a future (past) asymptotic state, and the point is consequently called a sink (source).  If the eigenvalues have both positive and negative real parts, then the equilibrium point is called  a saddle and is subsequently unstable to both the past and to the future.

%% -------------------------------------------------------------------------
%% --------------        Section 4                  ------------------------
%% -------------------------------------------------------------------------
\section{Qualitative Analysis}

%% 88888888888888888888888888888888888888888888888888888888888888888888888888888888888888888888888888888
%%
%% 88888888888888888888888888888888888888888888888888888888888888888888888888888888888888888888888888888

\subsection{${\cal P} = {\cal P}_0$}

\subsubsection{Equations}
The dynamical system (\ref{ds2}) is invariant under the transformation $({\cal P},\Sigma) \to (-{\cal P},-\Sigma)$.  This implies that we can restrict ourselves to the scenario ${\cal P}_0\geq 0$ without loss of generality.  The required analysis becomes more manageable if we also redefine the parameter ${\cal P}_0 = f(P)=P(P^2-1)$ where $f:[1,\infty)\to [0,\infty)$ is a homeomorphism.  Equation (\ref{constraint2}) can be used to eliminate the variable $\Omega_{\cal U}$ from the dynamical system (\ref{ds2}) to yield the following three dimensional system
\begin{eqnarray}\label{ds3}
\Sigma' & = & \Sigma(q-2)+ P(P^2-1), \nonumber\\
\Omega' & = & \Omega[2q-(3\gamma-2)],\\
\Omega_{\lambda}' &=& \Omega_{\lambda} [2q-(6\gamma -2)],\nonumber
\end{eqnarray}
where
\begin{equation}
q=1+\Sigma^2+\frac{1}{2}(3\gamma-4)\Omega+\frac{1}{2}(6\gamma-4)\Omega_{\lambda}.
\label{q2}
\end{equation}

The equilibrium points and their local stability are discussed below.

\subsubsection{Equilibrium Points}

%% -----------------------------------------------------------------------------
%% ------------------------    Point 1    --------------------------------------
%% -----------------------------------------------------------------------------
\noindent {\bf Bianchi I perfect fluid model} (BI$_{pf}$): 
$$ \Sigma = \frac{2P(P^2-1)}{3(2-\gamma)}, \quad   
\Omega = 1 - \frac{8P^2(P^2-1)^2}{9(2-\gamma)^2(3\gamma-4)}, \quad 
\Omega_{\lambda} = 0,\quad
q=\frac{3\gamma-2}{2}.$$
This equilibrium point represents an anisotropic Bianchi I perfect fluid cosmology. Because $1 \leq P < \infty$, we observe that $\Sigma \geq 0$. 
This point is physical only when $\Omega \geq 0$, which implies that $P$ and
$\gamma$ must also satisfy
\begin{equation} \label{omeg1}
1 - \frac{8P^2(P^2-1)^2}{9(2-\gamma)^2(3\gamma-4)} \geq 0. 
\end{equation}
For $\gamma<4/3$ the constraint is satisfied for all values of $P$.  However, for $4/3<\gamma < 2$ the constraint is essentially an inverted cubic function of $P^2$ with up to three roots.  It can be shown that $P^2=1$ is a local max for all values of $\gamma$, and that the largest real root is $P^2=\bar P_1$ where
$$\bar P_1=\frac{1}{4}\left[-(3\gamma-8)+\sqrt{ (3\gamma-8)^2-36(\gamma-2)^2}\right].$$
  Therefore the constraint is satisfied if $P^2$ is less than the value of the largest root, i.e, $P^2<\bar P_1$.
In the limit as the parameter $P\to 1$ this equilibrium point becomes isotropic and represents a zero curvature perfect fluid Friedmann-Robertson-Walker cosmology.  The exact solution at this equilibrium point has $H(t)=(2/3\gamma) t^{-1}$.  
The eigenvalues of the linearization about this point are:
\begin{eqnarray*}
\lambda_{1} & = & -3\gamma,\\
\lambda_{2,3} & = &
\frac{1}{4}\bigg[(9\gamma-14)\pm\sqrt{(9\gamma-14)^2+24(2-\gamma)(3\gamma-4)\Big(1-\frac{8P^2(P^2-1)^2}{9(2-\gamma)^2(3\gamma-4)}\Big)}\bigg].
\end{eqnarray*}
For $1 \leq \gamma < \frac{4}{3}$ this point exists for all values of $P$ and is locally a sink, whence
\begin{equation} \label{omeg1_u}
\Omega_{\cal U} = \frac{4P^2(P^2-1)}{3(2-\gamma)(3\gamma-4)} \leq 0.
\end{equation}
Over the interval $\frac{4}{3} < \gamma < 2$ the equilibrium point
becomes a local saddle and $\Omega_{\cal U} \geq 0$ whenever the point
exists. The stability of this point is summarized in Table \ref{table1}.

%% -----------------------------------------------------------------------------
%% ------------------------    Point 2    --------------------------------------
%% -----------------------------------------------------------------------------
\noindent {\bf Bianchi I brane-world model} (BI$_{br}$):
$$\Sigma = -\frac{P(P^2-1)}{3(\gamma-1)},\quad 
\Omega = 0,\quad 
\Omega_{\lambda} = 1 -\frac{P^2(P^2-1)^2}{9(\gamma-1)^2(3\gamma-2)}, \quad
q=3\gamma-1.$$
This equilibrium point represents an anisotropic Bianchi I brane-world cosmology since $\Omega_{\lambda}\not = 0$. 
The point is physical only when $\Omega_{\lambda}
\geq 0$ and thus $P$ and $\gamma$ must satisfy
\begin{equation}
1 - \frac{P^2(P^2-1)^2}{9(\gamma-1)^2(3\gamma-2)} \geq 0. 
\end{equation}  This constraint is essentially an inverted cubic function of $P^2$ with up to three roots.  It can be shown that $P^2=1$ is a local max for all values of $\gamma$, and the largest real root is $P^2=3\gamma-2$.  
  Therefore the constraint is satisfied if $P^2$ is less than the value of the largest root, i.e, $P^2<3\gamma-2$.
In the limit as the parameter $P\to 1$ this equilibrium point becomes isotropic and represents the usual Binetruy et al. \cite{Binetruy:2000a}  Friedmann-Robertson-Walker brane-world cosmology. The exact solution at this equilibrium point has $H(t)=(1/3\gamma) t^{-1}$.  
The eigenvalues of the linearization about this equilibrium point are:
\begin{eqnarray*}
\lambda_{1} & = & 3\gamma,\\
\lambda_{2,3} & = &
\frac{1}{2}\bigg[(9\gamma-7)\pm\sqrt{(9\gamma-7)^2-24(\gamma-1)(3\gamma-2)\Big(1-\frac{P^2(P^2-1)^2}{9(\gamma-1)^2(3\gamma-2)}\Big)}\bigg].
\end{eqnarray*}
This point is a source on the interval $1 < \gamma \leq 2$ whenever it exists. We also note the value
\begin{equation}
\Omega_{\cal U} = -\frac{P^2(P^2-1)^2}{3(\gamma-1)(3\gamma-2)}\leq 0.
\end{equation}
The stability of this point is summarized in Table \ref{table1}.

%% -----------------------------------------------------------------------------
%% ------------------------    Point 3    --------------------------------------
%% -----------------------------------------------------------------------------
\noindent {\bf Kasner-like bulk dominated brane-world model} (K$_{\cal U}$):
$$\Sigma = -P, \quad 
\Omega = 0, \quad
\Omega_{\lambda} = 0,\quad 
q=1+P^2. $$
This equilibrium point represents a Bianchi I model containing no ordinary matter but does contain a form of dark matter that can be interpreted as radiation as a result of the bulk effect  ${\cal U}$.  In the limit as $P\to 1$ this point approaches a vacuum dominated Bianchi I model or Kasner model.  This point exists for all values of the parameters $P$ and $\gamma$.  The exact solution at this equilibrium point has $H(t)=(1/(2+P^2))t^{-1}$.
The eigenvalues of the linearization about this equilibrium point are:
\begin{equation*}
\lambda_{1} = 2P^2+4-3\gamma, \qquad \lambda_{2} = 2P^2+4-6\gamma,
\qquad \lambda_{3} = 3P^2-1.
\end{equation*}
This point is a source for $P^2 > {3\gamma-2}$, else it is a saddle.
At this equilibrium point
\begin{equation}
\Omega_{\cal U} = 1 - P^2
\end{equation}
and therefore $\Omega_{\cal U} \leq 0$.  The stability of this
point is summarized in Table \ref{table1}.

%% -----------------------------------------------------------------------------
%% ------------------------    Point 4    --------------------------------------
%% -----------------------------------------------------------------------------
\noindent {\bf Bianchi I bulk dominated brane-world model} ($^\epsilon$BI$_{\cal U}$):
$$\Sigma = \frac{P + \epsilon \sqrt{4-3P^2}}{2}, \quad
\Omega = 0, \quad
\Omega_{\lambda} =0, \quad
q=(4-P^2)+\epsilon P\sqrt{4-3P^2}.$$
There are two copies of the equilibrium point, signified by the
parameter $\epsilon$ which takes values $\pm 1$. These equilibrium points again represent anisotropic Bianchi I models containing no ordinary matter but contain a form of dark radiation ${\cal U}$ as a result of the bulk.  However, these points differ from K$_{\cal U}$ in that $^+$BI$_{\cal U}$ approaches the a general relativistic Bianchi I model containing radiation as $P\to 1$ while the point $^-$BI$_{\cal U}$ approaches an isotropic Robertson-Walker model containing radiation. This point exists only if $P$ lies in the interval $1 \leq P \leq \frac{2}{\sqrt{3}}$.  The exact solution at this equilibrium point has $H(t)=(1/(q+1))t^{-1}$.
 The eigenvalues of the linearization about this equilibrium point are:
\begin{eqnarray*}
\lambda_{1} & = & \frac{4-3P^2+3\epsilon P \sqrt{4-3P^2}}{2},\\
\lambda_{2} & = & 6-P^2+\epsilon P \sqrt{4-3P^2}-3\gamma,\\
\lambda_{3} & = & 6-P^2+\epsilon P \sqrt{4-3P^2}-6\gamma.
\end{eqnarray*}
 Whenever this point exists this point is a saddle.  The remaining variable
\begin{equation}
\Omega_{\cal U} = \frac{3P^4}{4}\geq 0.
\end{equation}
 The stability of these points is summarized in Table \ref{table1}.

\begin{table}
\caption{\label{table1} This table summarizes the local stability of the equilibrium points in the ${\cal P}={\cal P}_0\equiv P(P^2-1)$ case.  DNE=Does Not Exist}
\begin{tabular}{|c|c|c|c|c|} \hline
Label & Equilibrium Point $(\Sigma,\Omega,\Omega_{\lambda})$ & Range of $\gamma$ & Conditions on $P$ & Stability\\
\hline\hline 
  &      & $1 \leq \gamma < \frac{4}{3}$ 
          & $1\leq P^2$ & Sink\\
 BI$_{pf}$ & \Big($\frac{2P(P^2-1)}{3(2-\gamma)}, 1 - \frac{8P^2(P^2-1)^2}{9(2-\gamma)^2(3\gamma-4)}, 0$\Big)
    & $\frac{4}{3} < \gamma < 2$ & $1\leq P^2<\bar P_1$ & Saddle \\
 & & $\frac{4}{3} < \gamma < 2$ & $\bar P_1<P^2$ & DNE \\
\hline BI$_{br}$ & \Big($-\frac{P(P^2-1)}{3(\gamma-1)}, 0, 1 -\frac{P^2(P^2-1)^2}{9(\gamma-1)^2(3\gamma-2)}$\Big) 
                 & $1 < \gamma \leq 2$ 
                 & $P^2<3\gamma-2$ & Source\\
                 & & $1<\gamma \leq 2$ & $P^2>3\gamma-2$ & DNE \\
\hline
K$_{\cal U}$ & ($-P, 0, 0$) & $1 \leq \gamma \leq 2$ & ${3\gamma-2}<P^2$ & Source\\
${}$ & ${}$ & ${}$ & $1\leq P^2 < {3\gamma-2}$ & Saddle\\
\hline
$^+$BI$_{\cal U}$ & $\Bigl(\frac{1}{2}(P +\sqrt{4-3P^2}), 0, 0 \Bigr)$ & $1 \leq \gamma \leq 2$ & $1 < P^2 < \frac{4}{3}$ & Saddle\\
${}$ & ${}$ & $1\leq \gamma \leq 2$ & $\frac{4}{3}<P^2$ & DNE\\
\hline
 
                       &      & $1 \leq \gamma < \frac{4}{3} $ & $ 1\leq P^2 < \frac{4}{3}$ & Saddle\\
                  ${}$ & ${}$ & $\frac{4}{3} < \gamma <\frac{14}{9}$ & $1\leq P^2 < \bar P_1$ & Sink\\
$^-$BI$_{\cal U}$ & $\Bigl(\frac{1}{2}(P -\sqrt{4-3P^2}), 0, 0 \Bigr)$  & $\frac{4}{3} < \gamma < \frac{14}{9}$ & $\bar P_1< P^2 <\frac{4}{3} $ & Saddle\\
                  ${}$ & ${}$ & $\frac{14}{9}<\gamma \leq 2$ & $ 1\leq P^2 < \frac{4}{3}$ & Sink\\
                  ${}$ & ${}$ & $1\leq\gamma \leq 2$ & $ \frac{4}{3}<P^2  $ & DNE\\
\hline
\end{tabular}\linebreak
where $\bar P_1=\frac{1}{4}\left((8-3\gamma)+\sqrt{(20-9\gamma)(3\gamma-4)}\right)$\hfill\hfill\linebreak
\end{table}

\subsubsection{Observations}

We observe that in general the dynamics in the case of ${\cal P}={\cal P}_0$ a constant, are quite different than what is found currently in the literature.  However, as we have observed in the preceding analysis, the qualitative behaviour approaches the qualitative behaviour discovered by Campos and Sopuerta \cite{Campos:2001a,Campos:2001b} in their analysis of the Bianchi I brane-worlds containing a perfect fluid.  In our analysis, for $P\not=1$ (${\cal P}_0\not = 0$) we have found that none of the equilibrium points are isotropic, and all have some form of dark matter in the form of radiation, that is ${\cal U}\not = 0$.  It appears that the non-local bulk corrections have a significant impact on the behaviour.  The initial singularity represented by either BI$_{br}$ and K$_{\cal U}$ are not isotropic.  This result suggests that the initial singularity in braneworld cosmologies is not isotropic as argued by Coley in \cite{Coley:2002a}.  As stated earlier, at each of the equilibrium points here, we have ${\cal U}={\cal U}_0t^{-2}$ and ${\cal P}_{\mu\nu}=[{\cal P}_{\mu\nu}]_0 t^{-2}$ where ${\cal U}_0$ is a constant and $[{\cal P}_{\mu\nu}]_0$ is a matrix of constants.

Of course the ansatz that $${\cal P}=\frac{6}{\kappa^2\lambda} \frac{\frac{1}{2}\sigma_{\mu\nu}{\cal P}^{\mu\nu}}      {\sqrt{3} H^2\sqrt{\frac{1}{2}\sigma_{\mu\nu}\sigma^{\mu\nu}}}={\cal P}_0$$ a constant may not be physically plausible,  however, it is no less acceptable than the alternative of artificially setting ${\cal P}_{\mu\nu}=0$ or ${\cal P}_{\mu\nu}\sigma^{\mu\nu}=0$ apriori.  In the next two subsections we consider $\cal P$ to be dynamic and analyze two alternative possibilities.

%% 88888888888888888888888888888888888888888888888888888888888888888888888888888888888888888888888888888
%%                          CASE 2
%% 88888888888888888888888888888888888888888888888888888888888888888888888888888888888888888888888888888

\subsection{${\cal P} = {\cal P}_{\Sigma}\Sigma^\alpha$,  $\alpha>0$}

\subsubsection{Equations}
To illustrate this case we let $\alpha = 1$ and again use equation
(\ref{constraint2}) to eliminate the variable $\Omega_{\cal U}$
from the dynamical system (\ref{ds2}) whence we obtain the following
three dimensional system
\begin{eqnarray}
\Sigma' & = & \Sigma(q-2)+ {\cal P}_{\Sigma}\Sigma,\nonumber\\
\Omega' & = & \Omega[2q-(3\gamma-2)],\label{ds4}\\
\Omega_{\lambda}' &=& \Omega_{\lambda} [2q-(6\gamma -2)],\nonumber
\end{eqnarray}
where again
\begin{equation*}
q=1+\Sigma^2+\frac{1}{2}(3\gamma-4)\Omega+\frac{1}{2}(6\gamma-4)\Omega_{\lambda}
\end{equation*}
and ${\cal P}_{\Sigma} \in \mathbb{R}$. We note that $\Sigma = 0$ is an
invariant set and that the dynamical system (\ref{ds4}) is
invariant under the transformation $\Sigma \to -\Sigma$. Therefore
without loss of generality we restrict ourselves to the invariant set
$\Sigma \geq 0$.
The equilibrium points and their local stability are discussed below.

\subsubsection{Equilibrium Points}

%% -----------------------------------------------------------------------------
%% ------------------------    Point 1    --------------------------------------
%% -----------------------------------------------------------------------------
\noindent {\bf FRW perfect fluid model} (F$_{pf}$): 
$$\Sigma =0, \quad \Omega = 1, \quad \Omega_{\lambda} = 0,\quad q = \frac{3\gamma-2}{2}.$$
This equilibrium point represents a zero curvature isotropic Friedman-Robertson-Walker perfect fluid cosmology and in the context of brane-world cosmologies when extended to the bulk is represented by the solutions in \cite{Csaki:1999,Cline:1999}. 
The exact solution at this equilibrium point has $H(t)=(2/3\gamma) t^{-1}$.The eigenvalues of the linearization about this equilibrium point are:
\begin{equation*}
\lambda_{1} = -3\gamma, \qquad \lambda_{2} = 3\gamma - 4, \qquad
\lambda_{3} = {\cal P}_{\Sigma} + \frac{3}{2}(\gamma - 2).
\end{equation*}
This point is a sink if $1 \leq \gamma < \frac{4}{3}$ and ${\cal
P}_{\Sigma} < -\frac{3}{2}(\gamma -2)$, else this point is a
saddle.  At this point
$\Omega_{\cal U} = 0$ and the stability of this point is summarized
in Table \ref{table2}.

%% -----------------------------------------------------------------------------
%% ------------------------    Point 2    --------------------------------------
%% -----------------------------------------------------------------------------
\noindent {\bf  FRW brane-world model} (F$_{br}$): 
$$\Sigma = 0, \quad \Omega = 0, \quad \Omega_{\lambda} = 1,\quad q = 3\gamma-1.$$
This equilibrium point represents the Binetruy et al. \cite{Binetruy:2000a}  Friedmann-Robertson-Walker brane-world solution. The exact solution at this equilibrium point has $H(t)=(1/3\gamma) t^{-1}$.
The eigenvalues of the linearization about this equilibrium point are:
\begin{equation*}
\lambda_{1} = 3\gamma, \qquad \lambda_{2} = 6\gamma - 4 , \qquad
\lambda_{3} = {\cal P}_{\Sigma} + 3(\gamma - 1).
\end{equation*}
This point is a source for $1 \leq \gamma \leq 2$ and
${\cal P}_{\Sigma} > -3(\gamma-1)$, else this point is a saddle.  At this point $\Omega_{\cal U} = 0$ and the stability of this point is summarized in Table \ref{table2}.

%% ----------------------------------------------------------------------------- 
%% ------------------------    Point 3    --------------------------------------
%% -----------------------------------------------------------------------------
\noindent {\bf  FRW bulk dominated brane-world model} (F$_{\cal U}$): 
$$\displaystyle\Sigma = 0, \quad \Omega = 0, \quad \Omega_{\lambda} = 0$$
This point represents a spatially homogeneous and isotropic cosmological model containing no ordinary matter but does contain a form of dark matter in the form of radiation as a result of the bulk term ${\cal U}$.
The eigenvalues of the linearization about this equilibrium point are:
\begin{equation*}
\lambda_1 = {\cal P}_{\Sigma}-1, \qquad \lambda_2 = 4-3\gamma,
\qquad \lambda_3 = 4-6\gamma,
\end{equation*}
where
\begin{equation*}
\Omega_{\cal U} = 1.
\end{equation*}
Over the interval $\frac{4}{3} < \gamma \leq 2$ when ${\cal
P}_{\Sigma} < 1$ this point is a sink, else the point is a
saddle. The stability of this point is summarized in Table \ref{table2}.

%% ----------------------------------------------------------------------------- 
%% ------------------------    Point 4    --------------------------------------
%% -----------------------------------------------------------------------------
\noindent {\bf  Kasner-like bulk dominated brane-world model} (K$_{\cal U}$):
$$\Sigma = \sqrt{1-{\cal P}_{\Sigma}}, \quad \Omega = 0, \quad \Omega_{\lambda} =0,\quad q= 2-{\cal P}_{\Sigma}.$$
This equilibrium point represents a Bianchi I model containing no ordinary matter but does contain a form dark matter in the form of radiation as a result of the bulk term ${\cal U}$.  This point is physical only when ${\cal P}_{\Sigma} \leq 1$.  In the limit as ${\cal P}_\Sigma\to 0$ this point approaches a vacuum dominated Bianchi I model or Kasner model.
The eigenvalues of the linearization about this equilibrium point are:
\begin{equation*}
\lambda_{1} = 2-2{\cal P}_{\Sigma}, \qquad \lambda_{2} = 6-2{\cal
P}_{\Sigma}-3\gamma, \qquad \lambda_{3} = 6-2{\cal
P}_{\Sigma}-6\gamma
\end{equation*}
where $$\Omega_{\cal U}={\cal P}_{\Sigma}.$$
This point is a source whenever ${\cal P}_{\Sigma} <
-3(\gamma-1)$, else it is a saddle.   
 The stability of this point is summarized in Table \ref{table2}.

\begin{table}
\caption{\label{table2} This table summarizes the local stability of the equilibrium points in the ${\cal P}={\cal P}_\Sigma\Sigma$ case.  DNE=Does Not Exist}
\begin{tabular}{|c|c|c|c|c|}\hline
Label & Equilibrium Point $(\Sigma,\Omega,\Omega_{\lambda})$ & Range of  $\gamma$ & Conditions on ${\cal P}_{\Sigma}$ & Stability\\
\hline\hline
    ${}$ & ${}$        & $1 \leq \gamma < \frac{4}{3}$ & ${\cal P}_{\Sigma} < -\frac{3}{2}(\gamma-2)$ & Sink\\
F$_{pf}$ & $(0, 1, 0)$ & $1 \leq \gamma < \frac{4}{3}$ & ${\cal P}_{\Sigma} > -\frac{3}{2}(\gamma-2)$ & Saddle\\
    ${}$ & ${}$        & $\frac{4}{3} < \gamma \leq 2$ & ${\cal P}_{\Sigma} \in \mathbb{R}$ & Saddle\\
\hline
F$_{br}$ & $(0, 0, 1)$ & $1 \leq \gamma \leq 2$ & ${\cal P}_{\Sigma} < -3(\gamma-1)$ & Saddle\\
    ${}$ & ${}$        & ${}$                   & ${\cal P}_{\Sigma} > -3(\gamma-1)$ & Source\\
\hline
F$_{\cal U}$ & $(0, 0, 0)$ & $1 \leq \gamma < \frac{4}{3}$ & ${\cal P}_{\Sigma} \in \mathbb{R}$ & Saddle\\
        ${}$ & ${}$        & $\frac{4}{3} < \gamma \leq 2$ & ${\cal P}_{\Sigma} < 1$ & Sink\\
\hline
        ${}$ & ${}$                                &  & ${\cal P}_{\Sigma} < -3(\gamma-1)$ & Source\\
K$_{\cal U}$ &$(\sqrt{1-{\cal P}_{\Sigma}}, 0, 0)$ & $1\leq \gamma\leq 2$ & ${\cal P}_{\Sigma} > -3(\gamma-1)$                  & Saddle\\
${}$ & ${}$                                &   & ${\cal P}_{\Sigma} > 1$ & DNE\\
\hline
\end{tabular}
\end{table}

\subsubsection{Observations}

The most interesting result in this case is that the Binetruy et al. \cite{Binetruy:2000a}  Friedmann-Robertson-Walker brane-world solution may not represent the past asymptotic state.  In particular there are parameter values of $P_\Sigma$ such that the initial singularity is represented by the somewhat more conventional Kasner-like solution.  We also find these cosmologies do not necessarily evolve to a purely general relativistic cosmology since one of the local sinks is dominated by the bulk term $\cal U$.

%% 88888888888888888888888888888888888888888888888888888888888888888888888888888888888888888888888888888
%%                          CASE 3
%% 88888888888888888888888888888888888888888888888888888888888888888888888888888888888888888888888888888

\subsection{${\cal P} = {\cal P}_{\Omega_{\cal U}}\Omega_{\cal U}^\beta$, $\beta>0$}

\subsubsection{Equations}
Similar to the previous case we will set $\beta = 1$ for
illustrative purposes only and again use equation (\ref{constraint2}) to eliminate
the variable $\Omega_{\cal U}$ from the dynamical system
(\ref{ds2}). We thus obtain the following three dimensional system
\begin{eqnarray}\label{ds5}
\Sigma' & = & \Sigma(q-2)+ {\cal P}_{\Omega_{\cal U}}(1-\Sigma^2-\Omega-\Omega_{\lambda}),\nonumber\\
\Omega' & = & \Omega[2q-(3\gamma-2)],\\
\Omega_{\lambda}' &=& \Omega_{\lambda} [2q-(6\gamma -2)],\nonumber
\end{eqnarray}
where again
\begin{equation*}
q=1+\Sigma^2+\frac{1}{2}(3\gamma-4)\Omega+\frac{1}{2}(6\gamma-4)\Omega_{\lambda},
\label{q3}
\end{equation*}
and ${\cal P}_{\Omega_{\cal U}} \in \mathbb{R}$.
The equilibrium points and their local stability are discussed below.

\subsubsection{Equilibrium Points}

%% -----------------------------------------------------------------------------
%% ------------------------    Point 1    --------------------------------------
%% -----------------------------------------------------------------------------
\noindent {\bf FRW perfect fluid model} (F$_{pf}$): 
$$\Sigma =0, \quad \Omega = 1, \quad \Omega_{\lambda} = 0, \quad q = \frac{3\gamma-2}2 $$
This equilibrium point exists for all parameter values and represents a zero curvature isotropic Friedman-Robertson-Walker perfect fluid cosmology and in the context of brane-world cosmologies when extended to the bulk is represented by the solutions in \cite{Csaki:1999,Cline:1999}. 
The exact solution at this equilibrium point has $H(t)=(2/3\gamma) t^{-1}$.  The eigenvalues for this point are:
\begin{equation*}
\lambda_{1} = -3\gamma, \qquad \lambda_{2} = 3\gamma - 4, \qquad
\lambda_{3} = \frac{3}{2}\gamma - 3.
\end{equation*}
This point is a sink over the interval $1 \leq \gamma < \frac{4}{3}$, else it is a saddle.  
At this point $\Omega_{\cal U} = 0$. The stability of this point
is summarized in Table \ref{table3}.

%% -----------------------------------------------------------------------------
%% ------------------------    Point 2    --------------------------------------
%% -----------------------------------------------------------------------------
\noindent {\bf  FRW brane-world model} (F$_{br}$): 
$$\Sigma = 0, \quad \Omega = 0, \quad \Omega_{\lambda} = 1,\quad q = 3\gamma-1.$$
This equilibrium point also exists for all parameter values and represents the the Binetruy-Deffayet-Langlois zero curvature brane-world solution \cite{Binetruy:2000a}.
The exact solution at this equilibrium point has $H(t)=(1/3\gamma) t^{-1}$.
The eigenvalues for this point are:
\begin{equation*}
\lambda_{1} = 3\gamma, \qquad \lambda_{2} = 6\gamma - 4, \qquad
\lambda_{3} = 3\gamma - 3.
\end{equation*}
This point is a source over the interval $1 < \gamma \leq 2$,
where again $\Omega_{\cal U} = 0$. The stability of this point is
summarized in Table \ref{table3}.

%% -----------------------------------------------------------------------------
%% ------------------------    Point 3    --------------------------------------
%% -----------------------------------------------------------------------------
\noindent {\bf  Bianchi I vacuum model (Kasner)} ($^\epsilon$K): 
$$\Sigma = \epsilon, \quad \Omega = 0, \quad \Omega_{\lambda} =0, \quad q=2.$$
The parameter $\epsilon = \pm 1$ signifying that there are two of these points that exist for all values of the parameters $\gamma,{\cal P}_{\Omega_{\cal U}}$.  These equilibrium points represent anisotropic vacuum Bianchi I models, known as Kasner cosmological models. The exact solution at this equilibrium point has $H(t)=(1/3) t^{-1}$. The eigenvalues for these point are:
\begin{equation*}
\lambda_{1} = 2 - 2\epsilon{\cal P}_{\Omega_{\cal U}}, \qquad
\lambda_{2} = 6-3\gamma, \qquad \lambda_{3} = 6-6\gamma.
\end{equation*}
This point is a saddle over the interval $1 < \gamma <2$,
regardless of the value of ${\cal P}_{\Omega_{\cal U}}$.  At this
equilibrium point ${\Omega_{\cal U}}=0$. The stability of this point is
summarized in Table \ref{table3}.

% -----------------------------------------------------------------------------
%% ------------------------    Point 4    --------------------------------------
%% -----------------------------------------------------------------------------
\noindent {\bf Kasner-like bulk dominated brane-world model} (K$_{\cal U}$): 
$$\Sigma ={\cal P}_{\Omega_{\cal U}}, \quad \Omega = 0,\quad \Omega_{\lambda} = 0, \quad q = 1+{\cal P}_{\Omega_{\cal U}}^2.$$
This equilibrium point represents a Bianchi I model containing no ordinary matter but does contain a form of dark radiation ${\cal U}$ as a result of the bulk.  
The exact solution at this equilibrium point has $H(t)=(1/{\cal P}_{\Omega_{\cal U}}^2) t^{-1}$.  The eigenvalues for this point are:
\begin{equation*}
\lambda_{1} = {\cal P}_{\Omega_{\cal U}}^2 - 1, \qquad
\lambda_{2} = 2{\cal P}_{\Omega_{\cal U}}^2 + 4 - 3\gamma, \qquad
\lambda_{3} = 2{\cal P}_{\Omega_{\cal U}}^2 + 4 - 6\gamma.
\end{equation*}
This point is a source over the interval $1 \leq \gamma \leq 2$,
given that ${\cal P}_{\Omega_{\cal U}}^2 > 3\gamma - 2$. 
 On the interval $\frac{4}{3} < \gamma < 2$ and given
that ${\cal P}_{\Omega_{\cal U}}^2 < \frac{3}{2}\gamma - 2$ this
point behaves as a sink.  For any other parameter values this point is a saddle. At this point
\begin{equation}
\Omega_{\cal U} = 1 - {\cal P}_{\Omega_{\cal U}}^2.
\end{equation}
The stability of this point is summarized in Table \ref{table3}.

% -----------------------------------------------------------------------------
%% ------------------------    Point 5    --------------------------------------
%% -----------------------------------------------------------------------------
\noindent {\bf Bianchi I perfect fluid model} (BI$_{pf}$): 
$$\Sigma = \frac{3\gamma-4}{2{\cal P}_{\Omega_{\cal U}}}, \quad \Omega = 1-\frac{3\gamma-4}{2{\cal P}_{\Omega_{\cal U}}^2}, \quad
\Omega_{\lambda} = 0, \quad q = \frac{3\gamma-2}{2} .$$
This equilibrium point represents an anisotropic Bianchi I perfect fluid cosmology. This point is physical when $\Omega \geq 0$, which implies that
${\cal P}_{\Omega_{\cal U}}$ and $\gamma$ must satisfy
\begin{equation}
1-\frac{3\gamma-4}{2{\cal P}_{\Omega_{\cal U}}^2} \geq 0.
\end{equation}
The eigenvalues for this point are:
\begin{eqnarray*}
\lambda_{1} &=& -3\gamma, \\
\lambda_{2,3} &=& \frac{1}{4}\left(3(\gamma-2)\pm\sqrt{[3(\gamma-2)]^2+24(3\gamma-4)(\gamma-2)\left(1-\frac{(3\gamma-4)}{2{\cal
P}_{\Omega_{\cal U}}^2}\right)}\right),
\end{eqnarray*}
where
\begin{equation}
\Omega_{\cal U} = \frac{3(3\gamma -4)(2-\gamma)}{4{\cal P}_{\Omega_{\cal
U}}^2}.
\end{equation}
This point, whenever it exists is a saddle over the interval $1 \leq \gamma <
\frac{4}{3}$, and it is a sink on the interval $\frac{4}{3} < \gamma < 2$.
 The stability of this point is summarized in Table \ref{table3}.

% -----------------------------------------------------------------------------
%% ------------------------    Point 6    --------------------------------------
%% -----------------------------------------------------------------------------
\noindent {\bf Bianchi I brane-world model} (BI$_{br}$):
 $$\Sigma = \frac{3\gamma-2}{{\cal P}_{\Omega_{\cal U}}}, \quad \Omega = 0, \quad 
\Omega_{\lambda} = 1-\frac{3\gamma-2}{{\cal P}_{\Omega_{\cal U}}^2},\quad q= 3\gamma-1 $$
This equilibrium point represents an anisotropic Bianchi I brane-world cosmology since $\Omega_{\lambda}\not = 0$. This point is physical only when $\Omega_{\lambda}
\geq 0$ and thus ${\cal P}_{\Omega_{\cal U}}$ and $\gamma$ must
satisfy
\begin{equation}
1-\frac{3\gamma-2}{{\cal P}_{\Omega_{\cal U}}^2} \geq 0.
\end{equation}
The eigenvalues for this point are:
\begin{eqnarray*}
\lambda_1 &=& 3\gamma, \\
\lambda_{2,3} &=&
\frac{1}{2}\left(3(\gamma-1)\pm\sqrt{[3(\gamma-1)]^2+24(3\gamma-2)(\gamma-1)\left(1-\frac{(3\gamma-2)}{{\cal
P}_{\Omega_{\cal U}}^2}\right)}\right),
\end{eqnarray*}
where
\begin{equation}
\Omega_{\cal U} = -\frac{3(3\gamma-2)(\gamma-1)}{{\cal P}_{\Omega_{\cal
U}}^2}.
\end{equation}
This point, whenever it exists, is a saddle over the interval $1 < \gamma \leq 2$, its
stability is independent of the parameter ${\cal P}_{\Omega_{\cal
U}}$. The stability of this point is summarized in Table \ref{table3}.

\begin{table}
\caption{\label{table3} This table summarizes the local stability of the equilibrium points in the ${\cal P}={\cal P}_{\Omega_{\cal U}}\Omega_{\cal U}$ case.  DNE=Does Not Exist}
\begin{tabular}{|c|c|c|c|c|}\hline
Label & Equilibrium Point $
(\Sigma,\Omega,\Omega_{\lambda})$ & Range of $ \gamma$ & Conditions on ${\cal P}_{\Omega_{\cal U}}^2$ & Stability\\
\hline\hline
F$_{pf}$     & $(0, 1, 0)$ & $1 \leq \gamma < \frac{4}{3}$ & ${\cal P}_{\Omega_{\cal U}}^2 \in \mathbb{R}^+$ & Sink\\
             &             & $\frac{4}{3} < \gamma < 2$    & ${\cal P}_{\Omega_{\cal U}}^2 \in \mathbb{R}^+$ & Saddle\\
\hline
F$_{br}$     & $(0, 0, 1)$ & $1 < \gamma \leq 2$ & ${\cal P}_{\Omega_{\cal U}}^2 \in \mathbb{R}^+$ & Source\\
\hline 
$^\epsilon$K & $(\epsilon, 0, 0)$                     & $1 < \gamma < 2$ & ${\cal P}_{\Omega_{\cal U}}^2 \in \mathbb{R}^+$ & Saddle\\
\hline
             &                                        & $1 \leq \gamma \leq 2$ & ${\cal P}_{\Omega_{\cal U}}^2 > 3\gamma - 2$ & Source\\
K$_{\cal U}$ & $({\cal P}_{\Omega_{\cal U}},0,0)$    & $1 \leq \gamma <\frac{4}{3}$ & ${\cal P}_{\Omega_{\cal U}}^2 < 3\gamma - 2$ & Saddle\\
             &                                        & $\frac{4}{3} < \gamma < 2$ & $\frac{3}{2}\gamma-2 < {\cal P}_{\Omega_{\cal U}}^2 < 3\gamma - 2$ & Saddle\\
             &                                        & $\frac{4}{3} < \gamma < 2$ & ${\cal P}_{\Omega_{\cal U}}^2 < \frac{3}{2}\gamma-2$ & Sink\\
\hline
             &                                        & $1 \leq \gamma <\frac{4}{3}$ & ${\cal P}_{\Omega_{\cal U}}^2 \in \mathbb{R}^+ $ & Saddle\\
BI$_{pf}$    & $\Big(\frac{3\gamma-4}{{\cal P}_{\Omega_{\cal U}}}, 1-\frac{3\gamma-4}{2{\cal P}_{\Omega_{\cal U}}^2}, 0\Big)$ 
                                                      & $\frac{4}{3} < \gamma <2$ & ${\cal P}_{\Omega_{\cal U}}^2 > \frac{3}{2}\gamma-2$ & Sink\\
             &                                        & $\frac{4}{3} < \gamma < 2$ & ${\cal P}_{\Omega_{\cal U}}^2 < \frac{3}{2}\gamma-2$ & DNE\\
\hline

BI$_{br}$ & $\Big(\frac{3\gamma-2}{{\cal P}_{\Omega_{\cal U}}}, 0, 1-\frac{3\gamma-2}{{\cal P}_{\Omega_{\cal U}}^2}\Big)$ 
                                                      & $1 < \gamma \leq 2$ & ${\cal P}_{\Omega_{\cal U}}^2 > 3\gamma-2$ & Saddle \\
          &                                           & $1 < \gamma \leq 2$ & ${\cal P}_{\Omega_{\cal U}}^2 < 3\gamma-2$ & DNE \\
\hline
\end{tabular}
\end{table}

\subsubsection{Observations}
An interesting observation in this case is that the Binetruy-Deffayet-Langlois Braneworld solution is always a source.  However, there will exist cosmological models that will asymptote to the past towards the anisotropic bulk-dominated Kasner-like models represented by the points ${\cal K}_{\cal U}$ when ${\cal P}_{\Omega_{\cal U}}>3\gamma-2$.  This means that the initial singularity is again not necessarily isotropic.  The future behaviour is determined primarily by the parameter $\gamma$.  If $\gamma<4/3$ then the general relativistic flat Robertson-Walker perfect fluid model is the future attractor.  However, if $\gamma>4/3$ then the non-local bulk corrections become dominant, that is, there exists models in which $\Omega_{\cal U}\not\to 0$ as $\tau\to\infty$.  Depending on the parameter ${\cal P}_{\Omega_{\cal U}}$ either ${\cal K}_{\cal U}$ or $BI_{pf}$ becomes the future asymptotic state.  It is worth noting that neither of these two local sinks are isotropic.

%% -------------------------------------------------------------------------
%% --------------            Conclusions            ------------------------
%% -------------------------------------------------------------------------
\section{Conclusions}
  
A dimensionless variable $\cal P$ was introduced to model the non-local bulk corrections arising from the bulk term ${\cal P}_{\mu\nu}$. That is, we introduced the quantity
\begin{equation}
{{\cal P}}= \frac{\sqrt{3}}{\kappa^2\lambda} \frac{\sigma_{\mu\nu}{\cal P}^{\mu\nu}}{\sigma H^2}.
\end{equation}
to model the non-local bulk corrections.  Since there are no equations in the brane-world approach to model the effects of the non-local bulk term ${\cal P}_{\mu\nu}$, there is no equation to determine the effects of ${\cal P}$.  Therefore, one is required to make justifiable and reasonable assumptions on the form of ${\cal P}$.   Here, we assume that ${\cal P}$ is a differentiable function of the remaining dynamical variables which results in self-similar asymptotic solutions.  

In this paper we specifically investigated three different cases, ${\cal P}={\cal P}_0$, ${\cal P}={\cal P}_{\Sigma}\Sigma$, and ${\cal P}={\cal P}_{\Omega_{\cal U}}\Omega_{\cal U}$ where ${\cal P}_0,{\cal P}_\Sigma$ and ${\cal P}_{\Omega_{\cal U}}$ are constants.
It was found that in each case, the effects of the non-local bulk effect ${\cal P}_{\mu\nu}$ were significant.  In the standard cosmological brane-world situation, the Binetruy-Deffayet-Langlois zero curvature brane-world solution \cite{Binetruy:2000a} represents the 'generic' initial behaviour in brane-world cosmologies containing ordinary matter.  Here we find that Kasner-like bulk-dominated models can also act as sources in all three cases and that the Binetruy-Deffayet-Langlois zero curvature brane-world solution does not even exist as an equilibrium state in one of the cases studied.  With regards to possible future behaviors, we find for $1\leq \gamma < 4/3$ that the future asymptotic state is generically matter dominated, in that the future asymptotic state has $\Omega_{pf}\not\to 0$ as $\tau\to\infty$.  For $\gamma>4/3$ we find that there exist future asymptotic states that can become 'bulk-dominated' in that ${\Omega_{\cal U}}\not\to 0$ as $\tau \to \infty$.

 The qualitative analysis of this system is by no means complete.  Only a local analysis of finite equilibrium points was investigated, none of the global behaviour was determined and none of the bifurcations were analyzed in any detail. There may be some additional interesting dynamics to be found, at points of infinity or even closed/periodic orbits at finite values. Further work in this area should include a discussion of these topics, but the motivation for this paper was to illustrate the effects of the non-zero non-local bulk effect ${\cal P}_{\mu\nu}$ in which one makes the assumption that all equilibrium states of the resulting dynamical system represent self-similar cosmologies. 

In conclusion, it should be apparent that neglecting the bulk effects in the brane-world approach to brane-world cosmologies is extremely dangerous.  Even in this simple toy model, with a relatively simple ansatz for the non-local bulk term ${\cal P}_{\mu\nu}$, we find significant differences with what is currently found in the literature when one ignores these effects. 

%% -------------------------------------------------------------------------
%% --------------         Acknowledgments           ------------------------
%% -------------------------------------------------------------------------

\ack
 Both RvdH and AH are supported in part through a research grant from the Natural Sciences and Engineering Research Council of Canada and a research grant from the St. Francis Xavier University Council on Research.  RvdH would also like to thank A.A. Coley for comments on an earlier draft of this paper.

%% -------------------------------------------------------------------------
%% --------------            References             ------------------------
%% -------------------------------------------------------------------------

\end{document}